\documentstyle[preprint,epsfig,aps]{revtex}
\begin{document}
\draft
 
\title{ Energy Dependence of the  Near-Threshold 
        Total Cross-Section for the $pp\rightarrow pp\eta^{\prime}$ Reaction.
       }        
\author{P.~Moskal$^{1,2}$, 
        H.-H.~Adam$^3$,
        J.T.~Balewski$^{4,5}$,
        V.~Baru$^{2,6}$,
        A.~Budzanowski$^4$,
        D.~Grzonka$^2$,
        J.~Haidenbauer$^2$,
        L.~Jarczyk$^1$,
        A.~Khoukaz$^3$,
        K.~Kilian$^2$,
        M.~K\"{o}hler$^7$,
        P.~Kowina$^8$,
        A.~Kudryavtsev$^6$,
        N.~Lang$^3$,
        T.~Lister$^3$,
        W.~Oelert$^2$,
        C.~Quentmeier$^3$,
        R.~Santo$^3$,
        G.~Schepers$^{2,3}$,
        T.~Sefzick$^2$,
        S.~Sewerin$^2$, 
        M.~Siemaszko$^8$,
        J.~Smyrski$^1$,
        A.~Strza{\l}kowski$^1$,
        M.~Wolke$^2$,   
        P.~W{\"u}stner$^7$,
        W.~Zipper$^8$
       }
\address{$^1$ Institute of Physics, Jagellonian University, PL-30-059 Cracow, Poland}
\address{$^2$ IKP, Forschungszentrum J\"{u}lich, D-52425 J\"{u}lich, Germany}
\address{$^3$ IKP, Westf\"{a}lische Wilhelms--Universit\"{a}t, D-48149 M\"{u}nster, Germany}
\address{$^4$ Institute of Nuclear Physics, PL-31-342 Cracow, Poland}
\address{$^5$ IUCF, Bloomington, Indiana, IN-47405, USA}
\address{$^6$ Institute of Theoretical and Experimental Physics, 117258 Moscow, Russia}
\address{$^7$ ZEL,  Forschungszentrum J\"{u}lich, D-52425 J\"{u}lich,  Germany}
\address{$^8$ Institute of Physics, Silesian University, PL-40-007 Katowice, Poland} 

\date{\today}
\maketitle
\begin{abstract}
Total cross sections for the $pp \rightarrow pp\eta^{\prime}$ reaction
have been measured in the excess energy range from Q~=~1.53~MeV to 
Q~=~23.64~MeV.
The experiment has been performed at the internal installation COSY~-~11~\protect\cite{brau96}
using a stochastically cooled proton beam of the COoler  SYnchrotron COSY~\protect\cite{maie97}
and a hydrogen cluster target~\protect\cite{domb97,khou96}.
The determined energy dependence of the total cross section 
weakens the hypothesis of the S-wave repulsive interaction between 
the $\eta^{\prime}$ meson and the proton~\cite{baruej,moskalacta}.
New data agree well with predictions based on the 
phase-space distribution modified by the proton-proton 
final-state-interaction (FSI) only.
\end{abstract}
$~~$\\

Recently, total cross sections for the production of the $\eta^{\prime}$ meson
in the collision of protons close to the reaction threshold 
have been published~\cite{hiboupl,moskalprl} for the first time. 
Two independent experiments performed at the accelerators
SATURNE and COSY have delivered consistent results.
The data has triggered off an interest  in explaining the unknown
dynamics of the $pp \rightarrow pp\eta^{\prime}$ 
reaction~\cite{nakayama,bernard,gedalin,sibirtsev,wilk,sibi2}. 
The determined total cross sections are about a factor of thirty smaller 
than the ones for the $pp \rightarrow pp\eta$ 
reaction~\cite{hiboupl,caleneta,pinot,bergdolt,smyrskipl} 
at the corresponding values of excess energy. 
Trying to explain this large difference Hibou~et~al.~\cite{hiboupl}
showed that calculations within a one-pion exchange model, where the parameters 
were adjusted to fit the total cross section for the $pp \rightarrow pp\eta$ 
reaction, underestimate the $\eta^{\prime}$ cross sections by about a factor of 
two. This discrepancy suggests that short-range production mechanisms as for 
example heavy meson exchange, mesonic currents~\cite{nakayama}, or more exotic 
processes like the production via a fusion of  gluons~\cite{bass} may 
contribute significantly to the creation of $\eta$ and $\eta^{\prime}$ 
mesons~\cite{wilk}. Such effects are likely, since the momentum transfer 
required to create these mesons is much larger than for the pion production, 
and already in case of the $pp\rightarrow pp\pi^{0}$ reaction a significant 
short-range heavy meson exchange contribution is necessary in order to obtain
agreement with experimental results~\cite{haiden,horowitz}.
On the other hand, Sibirtsev and~Cassing~\cite{sibirtsev} concluded that the 
one-pion exchange model, including the proton-proton final state interaction 
(pp~-~FSI), reproduces the magnitude of the experimental data and hence, 
the other exchange currents either play no role or cancel each other.

It is well established that the $\eta$ meson is predominantly produced via the 
excitation of an intermediate baryonic 
resonance~S$_{11}$(1535)~\cite{caleneta,germondwilk,laget,vetter,moalem,faldtwilk1}.
Both, the large difference in the production cross sections for $\eta$ and $\eta^{\prime}$ mesons,
and the lack of experimentally established baryonic resonances, which would 
decay into $\eta^{\prime}$, suggest that the $pp\rightarrow pp\eta^{\prime}$ 
reaction occurs without an excitation of the colliding protons.
Indeed, as demonstrated by Gedalin~et~al.~\cite{gedalin}, the magnitude of the 
close-to-threshold $\eta^{\prime}$ production can be explained without a 
resonant production term. However, for the $\eta^{\prime}$-photoproduction off 
protons~\cite{ploetzke,abbhhm,ahhm} the excitation function   
is described by an assumed coherent excitation of two possible resonances~\cite{ploetzke} (S$_{11}$(1897) 
and P$_{11}$(1986)), which decay into $\eta^{\prime}$ and proton.
Anticipating this hypothesis, recently Nakayama~et~al.~\cite{nakayama}
have shown that it is also possible to explain the 
magnitude and energy dependence of the close-to-threshold total cross section for the 
$pp\rightarrow pp\eta^{\prime}$
reaction assuming a dominance of these resonances and choosing an appropriate ratio
of pseudoscalar to pseudovector coupling.  However, the mesonic and nucleonic currents alone can describe
the data as well~\cite{nakayama}. 

The ambiguities in the description of the $pp\rightarrow pp\eta^{\prime}$
reaction mechanisms, which are partly due to the poorly known coupling constants,
indicate that the theory of the $\eta^{\prime}$ meson creation is still far from delivering 
a complete and  univocal picture of the process and call for further theoretical and experimental effort.
A possible gluonium admixture in the $\eta^{\prime}$ meson makes the study
even more complicated but certainly also more interesting.
Albeit the quark content of $\eta$ and $\eta^{\prime}$ mesons is very similar
a fusion of gluons emitted from the exchanged
quarks of the colliding protons~\cite{kolacosynews} would  contribute primarily
to the creation of the $\eta^{\prime}$ meson which is predominantly a flavour singlet
state due to the small  pseudoscalar mixing angle
($\Theta_{PS} \approx -15^{o}$)~\protect\cite{bram97}. 
 
Another complication in understanding the production mechanism
is the unknown $\eta^{\prime}$-proton interaction, 
which is of course in itself an interesting issue to be studied.
One of the remarkable features of the published results on $\eta$ and $\eta^{\prime}$ production is 
that the energy dependence of the total cross section appears not to follow the predictions based 
on the phase-space volume folded by the proton-proton
final state interaction, which is the case in the $\pi^{0}$ meson production~\cite{meyer90,meyer92}.
Moreover, for $\eta$ and $\eta^{\prime}$ mesons
the deviations from such predictions were qualitatively different:
The close-to-threshold cross sections for the $\eta$ meson
are strongly enhanced compared to the model comprising only the proton-proton interaction~\cite{caleneta},
opposite to the observed suppression in case of the $\eta^{\prime}$~\cite{baruej,moskalacta}.
The energy dependence of the total cross section for the $pp \rightarrow pp\eta$ reaction
can be described when  the $\eta$-proton attractive interaction is taken into
account~\cite{ulf95,moalem1}.
Although the $\eta$-proton interaction is much weaker than the proton-proton one
(compare the scattering length a$_{p\eta} = 0.751$~fm + $i$~0.274~fm~\cite{greenwycech}  with a$_{pp} =
-7.83$~fm~\cite{naisse})
it becomes important through the interference terms between the various final pair interactions~\cite{moalem1}.
  By analogy, the steep decrease of the total cross section when approaching the kinematical threshold
for the $pp \rightarrow pp\eta^{\prime}$ reaction
could have been explained assuming a repulsive
$\eta^{\prime}$-proton interaction~\cite{baruej,moskalacta}.
This interpretation, however, should rather be excluded
now in view of the new COSY~-~11 data reported in this letter.

The experiment has been performed at the cooler synchrotron COSY-J{\"u}lich~\cite{maie97},
using the COSY~-~11 facility~\cite{brau96} and the $H_{2}$ cluster
target~\cite{domb97,khou96} installed in front of one of the regular COSY dipole magnets.
The target, which is realized as a beam of $H_{2}$ molecules grouped to clusters of up to 
$10^{6}$ atoms, crosses perpendicularly the beam of $\sim 2\cdot 10^{10}$ protons circulating 
in the ring. The beam of accelerated protons is cooled stochastically during the measurement cycle. 
Longitudinal and vertical cooling enables to keep the circulating beam 
practically without energy losses and without a spread of its dimensions
when passing  $~1.6\cdot 10^{6}$ times per second through the 
10$^{14}$~atoms/cm$^{2}$ thick target 
during a 60~minutes cycle. The beam dimensions are determined from the distribution
of elastically scattered protons and are found to be 2~mm and 5~mm in the 
horizontal and vertical direction, respectively~\cite{moskalnewrep}. Quoted values denote standard deviations
of an assumed Gaussian beam density distribution.
The $pp \rightarrow pp\eta^{\prime}$ reaction has been investigated at eight different energies of a 
proton beam corresponding to excess energies ranging from Q~=~1.53~MeV to Q~=~23.64~MeV
as listed in table~\ref{cross_etap}. The total integrated luminosity obtained during two weeks of the 
experiment amounts to~1.4~pb$^{-1}$, and was monitored by the simultaneous measurement 
of elastically scattered protons. A comparison of the measured differential distributions with 
results from the literature~\cite{albers} determines the absolute luminosity with the statistical 
accuracy of 2.5~$\%$ for each excess energy.

\begin{figure}[H]
 \unitlength 1.0cm
  \begin{picture}(12.2,17.0)
    \put(2.0,7.7){
       \epsfig{figure=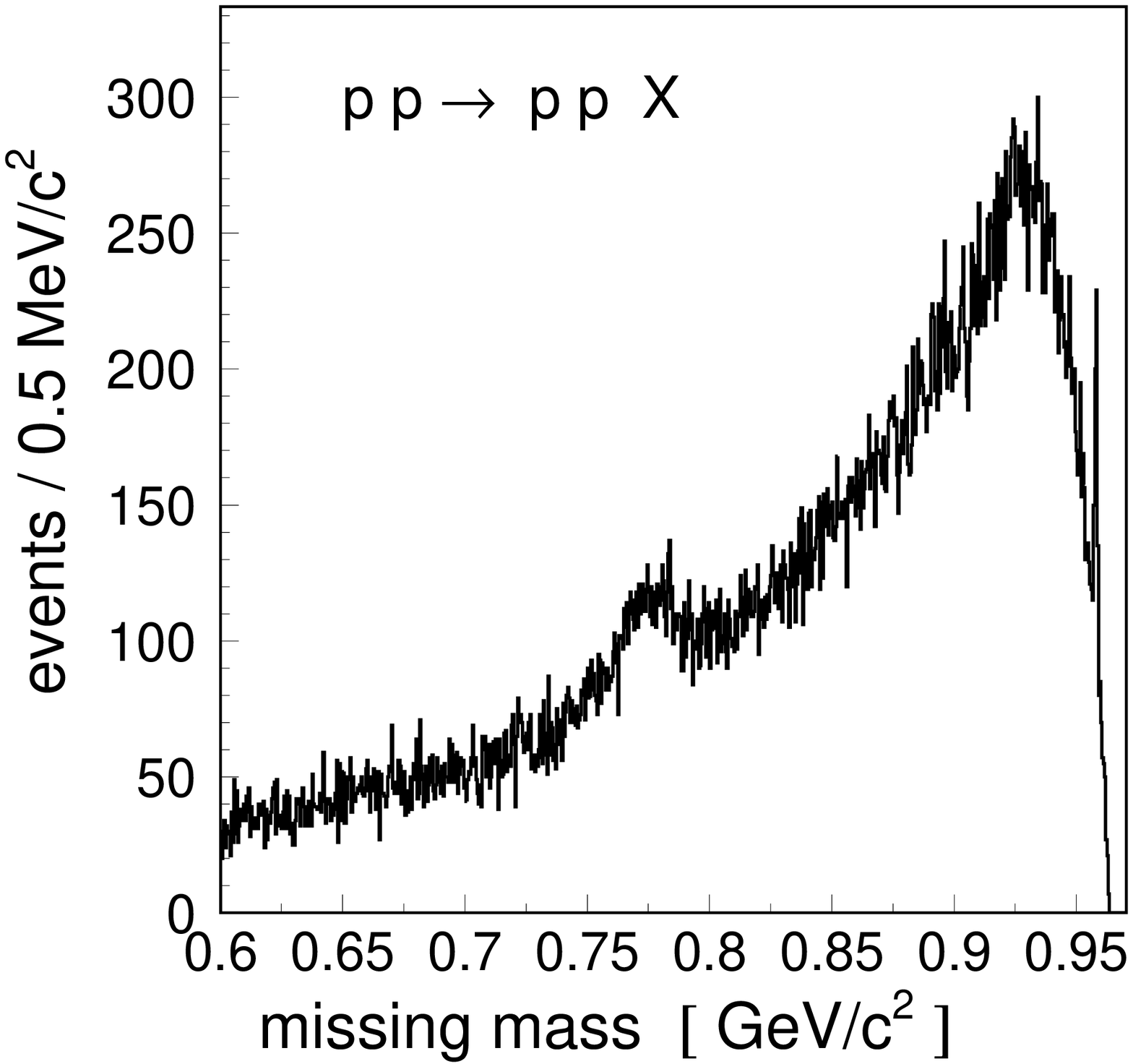,height=7.3cm,width=9.0cm,angle=0}
    }
       \put(10.0,7.7){
          { a)}
       }
    \put(2.0,0.0){
       \epsfig{figure=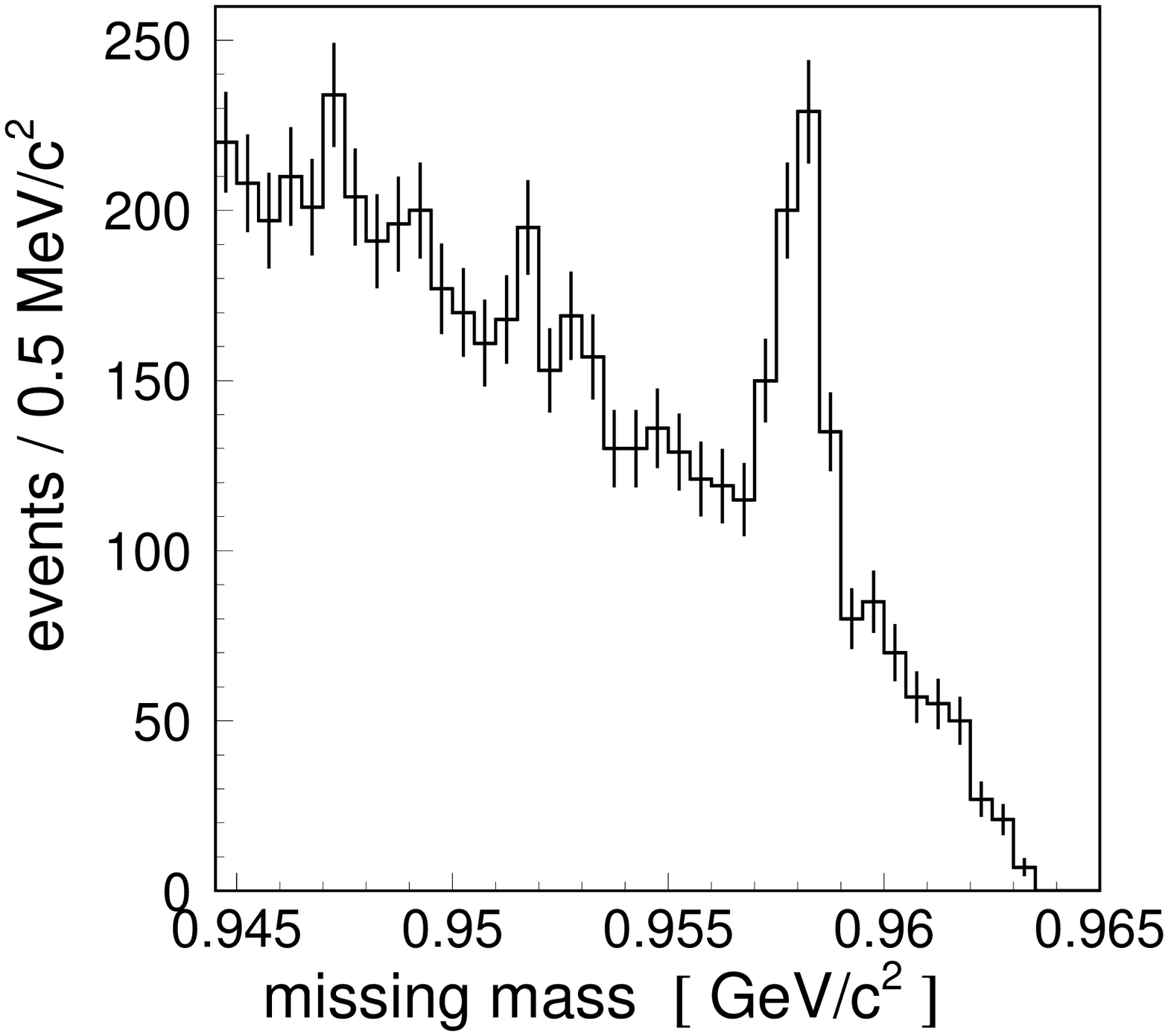,height=7.3cm,width=9.0cm,angle=0}
    }
       \put(10.0,0.0){
          { b)}
       }
  \end{picture}
  \vspace{2.0cm}
  \caption{
           Mass spectrum of the unobserved particle or system of particles
           in the $pp\rightarrow ppX$ reaction
           determined at Q~=~5.8~MeV above the $\eta^{\prime}$ production threshold.
        }
\label{miss1}
\end{figure}

\begin{figure}[H]
 \unitlength 1.0cm
  \begin{picture}(12.2,18.0)
    \put(3.0,0.0){
       \epsfig{figure=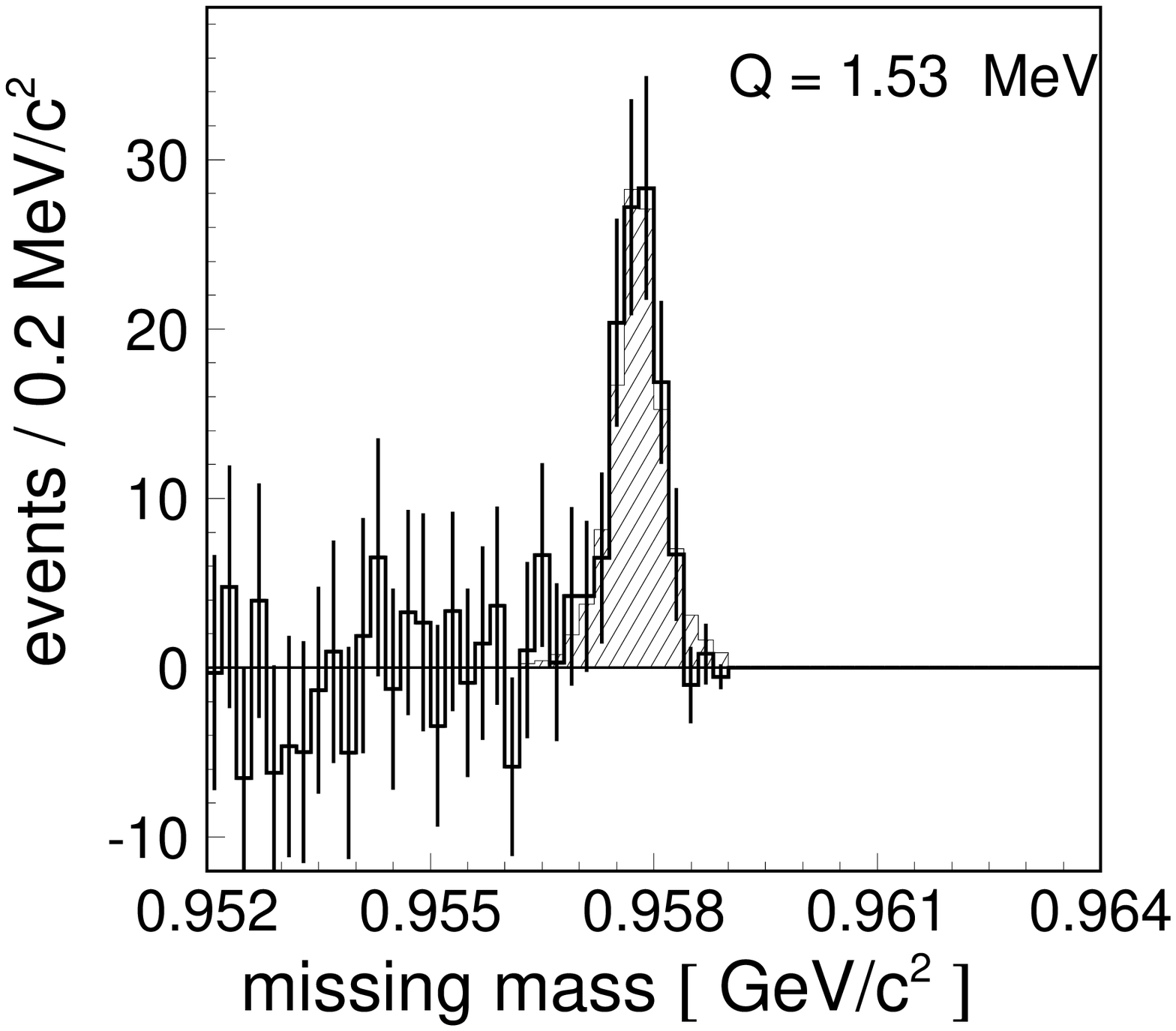,height=5.2cm,width=7.0cm,angle=0}
    }
       \put(9.0,0.0){
          { d)}
       }
    \put(3.0,4.5){
       \epsfig{figure=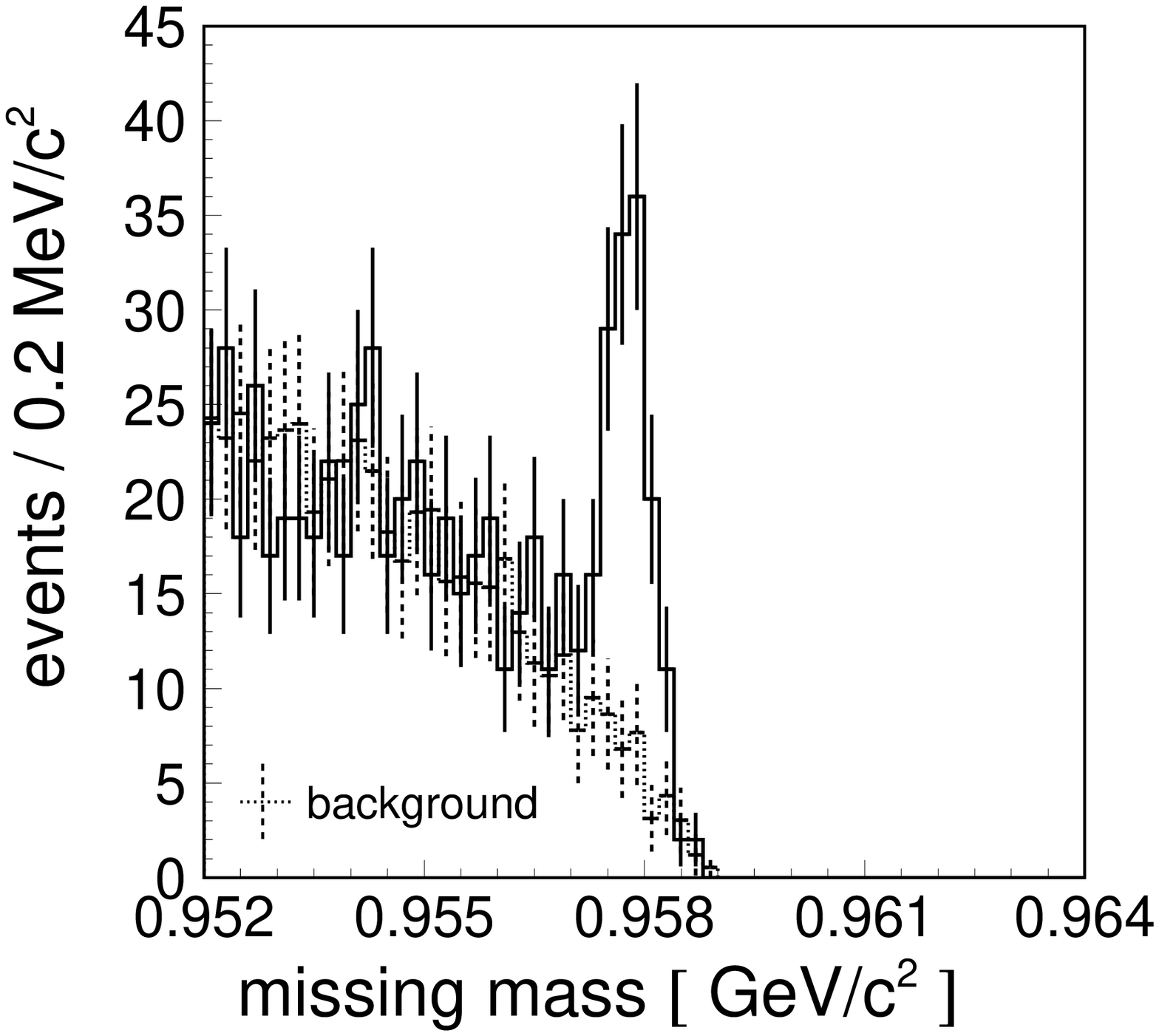,height=5.2cm,width=7.0cm,angle=0}
    }
       \put(9.0,4.5){
          { c)}
       }
    \put(3.0,9.0){
       \epsfig{figure=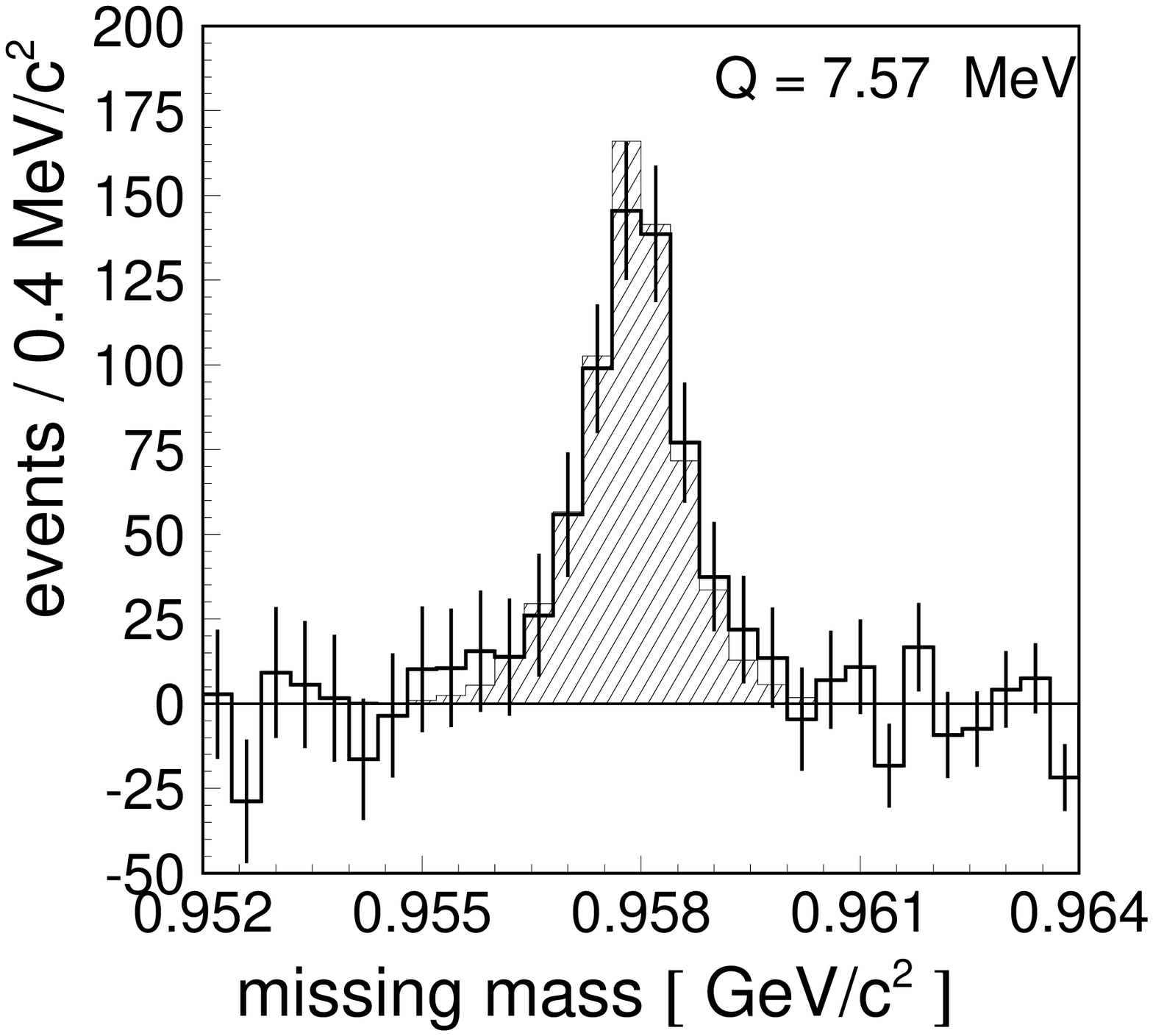,height=5.2cm,width=7.0cm,angle=0}
    }
       \put(9.0,9.0){
          { b)}
       }
    \put(3.0,13.5){
       \epsfig{figure=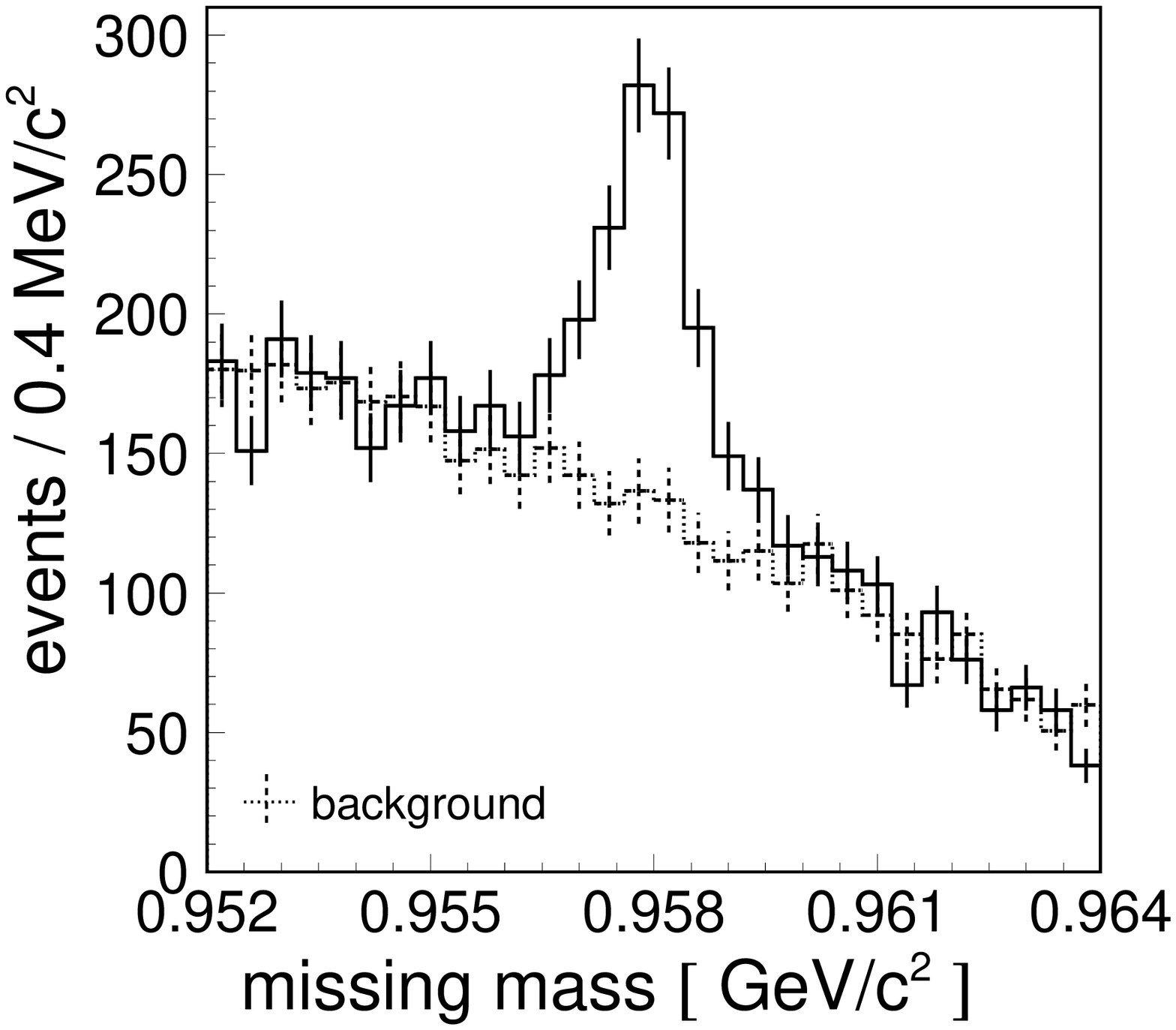,height=5.2cm,width=7.0cm,angle=0}
    }
       \put(9.0,13.5){
          { a)}
       }
  \end{picture}
  \vspace{1.4cm}
  \caption{
           Missing mass distribution with respect to the proton-proton system: \  \
           (a),(b)   measurements at Q~=~7.57~MeV 
           and  (c),(d) at Q~=~1.53~MeV.
           Background shown  as  dotted lines is combined from the measurements at different energies
           shifted to the appropriate kinematical limits and normalized to the solid-line histogram.
           Dashed histograms are obtained by means of Monte-Carlo simulations.
        }
\label{miss2}
\end{figure}

If at the intersection point of the cluster beam with  the COSY proton beam the collision of protons
results  in the production of a  meson, then
the ejected protons -~having smaller momenta than the beam protons~- are separated
from the circulating beam by the magnetic field.
Further they leave the vacuum chamber through a thin exit foil
and are registered by the detection system consisting of drift chambers and scintillation
counters~\cite{brau96,smyrskipl}.
The hardware trigger, based on signals from scintillation detectors, 
was adjusted to register all events with at least two positively charged
particles~\cite{moskalphd}. Tracing back trajectories from the drift chambers 
through the dipole magnetic field to the target point allowed for the determination of the particle momenta.
From momentum and velocity, the latter measured using scintillation detectors,
it is possible to identify the mass of the particle. Figure~\ref{invmass} 
shows the squared mass of two simultaneously detected particles.  
A~clear separation is seen into groups of events with two protons,
two pions, proton and pion and also deuteron and pion. This spectrum
enables to select events with two registered protons.
The knowledge of the momenta of both protons before and
after the reaction allows to calculate the mass of an unobserved particle or system of particles
created in the reaction. Figure~\ref{miss1}a depicts the missing mass spectrum obtained for the
$pp \rightarrow pp X$ reaction at an excess energy  of Q~=~5.8~MeV above the $\eta^{\prime}$
meson production threshold. Most of the entries in this spectrum originate 
from the multi-pion production~\cite{moskalprl,moskalphd}, forming a continuous background to the well 
distinguished peaks accounting for the creation of $\omega$ and $\eta^{\prime}$ mesons, which can be 
seen at  mass values of 782~MeV/c$^{2}$ and 958~MeV/c$^{2}$, respectively.
The signal of the $pp\rightarrow pp\eta^{\prime}$ reaction
is better to be seen in Figure~\ref{miss1}b, where the missing mass distribution in
the vicinity of its kinematical limit is presented. Figure~\ref{miss2}a shows the missing mass spectrum
for the measurement at Q~=~7.57~MeV together with the multi-pion background~(dotted line) as
combined  from the measurements at different excess energies~\cite{moskalnewrep}.
Subtraction of the background leads to the spectrum with a clear signal at the mass of the
$\eta^{\prime}$ meson as shown by the solid line in Figure~\ref{miss2}b. The dashed histogram
in this figure corresponds to Monte-Carlo simulations where the beam and target conditions
were deduced from the measurements of  elastically scattered protons~\cite{moskalnewrep}.
The magnitude of the simulated distribution was fitted to the data, but the consistency of the
widths is a measure of understanding of the detection system and the target-beam conditions.
Histograms from a measurement at Q~=~1.53~MeV shown in Figures~\ref{miss2}c,d demonstrate
the achieved missing-mass resolution at the COSY-11 detection system, when using a stochastically cooled proton beam.
The width of the missing mass distribution (Fig.~\ref{miss2}d), which is now close to the natural width
of the $\eta^{\prime}$ meson ($\Gamma_{\eta^{\prime}}=0.203$~MeV~\cite{pdb98}), 
is again well reproduced by the Monte-Carlo simulations.
The broadening of the width of the $\eta^{\prime}$ signal with increasing excess energy
(compare Figs.~\ref{miss2}b and~\ref{miss2}d) is a kinematical effect discussed in more
detail in reference~\cite{smyrskipl}. The decreasing signal-to-background ratio with growing excess energy 
is due to the broadening of the $\eta^{\prime}$ peak and the increasing  
background~(see Fig.~\ref{miss1}b) when moving away from the kinematical limit.
At the same time,  the shape of the background, determined by the convolution
of the detector acceptance and the distribution of the two- and three-pion production~\cite{moskalphd},
remains unchanged within the studied range of beam momenta from 3.213~GeV/c to 3.283~GeV/c.
The signal-to-background ratio changes from 1.8 at Q~=~1.53~MeV to 0.17 at Q~=~23.64~MeV.
The geometrical acceptance, being defined by the gap of the dipole magnet 
and the scintillation detector most distant from the target~\cite{brau96,smyrskipl},
decreases from 50~$\%$ to 4~$\%$ within this range of excess energies. However, 
in the horizontal plane the range of polar scattering angles is still unlimited.
The calculated acceptance depends on the angular distribution of the reaction products,
which was assumed to be defined by the three body phase-space and the 
interaction of the outgoing protons. Calculating acceptance, the proton-proton FSI was taken into account
by weighting phase-space generated events by the square of the proton-proton $^{1}$S$_{0}$-wave 
amplitude, $|$A$|$$^{2}$ . The enhancement, $|$A$|$$^{2}$, from the proton-proton FSI was 
estimated as an inverse of the squared Jost function,
with Coulomb interaction being taken into account~\cite{druzhinin}.
Generally, the attractive proton-proton FSI lowers the angle between outgoing
protons, increasing the acceptance. However, at the same time the efficiency
for the reconstruction of both proton trajectories decreases.
For the first five measurements denoted in table~\ref{cross_etap} both
effects are in the order of 3~$\%$ and cancel each other.
An increase of the overall efficiency is crucial only for the last two points 
listed in table~\ref{cross_etap} and amounts to 9~$\%$ and 25~$\%$ for
Q~=~14.21~MeV and Q~=~23.64~MeV, respectively. 
In order to estimate a systematical error due to the inaccuracy 
of the pp-FSI, we calculated the acceptance using another prescription 
for $|$A$|$$^{2}$, which was obtained from the
phase-shifts~\cite{morton} calculated according to the modified  Cini-Fubini-Stanghellini
formula with the Wong-Noyes Coulomb correction~\cite{naisse,noyeslip,noyes}.
Now the obtained efficiency was 13~$\%$ and 34~$\%$ larger as compared to the 
pure phase-space calculations for Q~=~14.21~MeV and Q~=~23.64~MeV, respectively.
Thus, for the highest energy there is a 9~$\%$ difference depending on the applied prescription.
The second main source of the systematical error is the inaccuracy 
of the determination of the two-track reconstruction efficiency.
This was established to be 9~$\%$ at Q~=~1.53~MeV~\cite{moskalphd} and close to zero at Q~=~23.64~MeV. 
This uncertainty decreases with increasing Q, 
since at higher excess energy the probability that the tracks of the protons will be too close 
to be unresolved by the drift chambers is reduced. 
In addition to the discussed  sources of the systematical error,
which add up to 9~$\%$ inaccuracy independent of energy,
there are further systematical uncertainties with respect to i) the geometry 
of the detection system (2~$\%$), ii) the estimated losses due to the multiple scattering 
or nuclear reactions~(1~$\%$) and iii) the luminosity determination~(3~$\%$)~\cite{moskalphd}.  
Hence, the overall systematical error of the cross section values,
including the normalization uncertainty,  amounts to 15~$\%$.

Figure~\ref{etap-fsi} shows the compilation of  total cross sections for the $\eta^{\prime}$
meson production. The data  reported here are shown as filled
circles. The absolute value of the excess energy was determined from the position
of the $\eta^{\prime}$ peak in the missing mass spectrum, which should correspond to
the mass of the meson $\eta^{\prime}$. The systematical error of the excess energy
established by this method equals to 0.44~MeV and constitutes of 0.14~MeV due to the uncertainty
of the $\eta^{\prime}$ meson mass~\cite{pdb98} and of 0.3~MeV due to the
inaccuracy of the detection system geometry~\cite{moskalabsolut}, with the largest effect
originating from the inexactness of relative settings of target, dipole and drift chambers. 

 The solid line depicts calculations of the total cross section assuming that the
primary production amplitude is constant and that only a proton-proton interaction significantly 
influences the exit channel. 
The magnitude was fitted to the data and the obtained $\chi^{2}$ value per degree of freedom
amounts to 1.6. An~inclusion of the $\eta^{\prime}$-proton interaction in the scattering length approximation,
by factorizing p-p and $\eta^{\prime}$-p FSI, resulted in a rather modest estimation 
of the real part of the $\eta^{\prime}$-proton scattering length: $|Re~a_{\eta^{\prime}p}|~<~0.8$~fm. 
The proton-proton scattering amplitude was computed according to the formulas from 
reference~\protect\cite{druzhinin}. The obtained energy dependence (solid line in Fig.~\ref{etap-fsi}) 
agrees within a few line thicknesses with the model developed by F{\"a}ldt and 
Wilkin~\protect\cite{faldtwilk}.
  
The  present data show that the phase-space volume weighted by the proton-proton FSI describes the 
near-threshold energy dependence of the total cross section for the $pp \rightarrow pp\eta^{\prime}$ 
reaction quite well. The influence of the $\eta^{\prime}$-proton FSI on the energy dependence of the 
total cross section  is too weak to be seen within the up-to-date experimental accuracy. 
Based on the energy dependence of the total cross section only, it is impossible to 
decouple effects from $\eta^{\prime}$-proton FSI and primary production amplitude.
As shown by Nakayama~et~al.~\cite{nakayama} the variation of the energy dependence of the total cross 
section, due to the production mechanism in the discussed energy range, can be in the order of 10~$\%$.
To learn more about the $\eta^{\prime}$-proton interaction a determination of differential cross
sections is required. 

It is interesting to note that in proton-proton collisions at much higher momenta~(450~GeV/c) 
the $\eta$ and $\eta^{\prime}$ mesons seem to have a similar production mechanism which differs from that of the~$\pi^{0}$
one~\cite{barberis}. However, close to threshold the data show similarities between $\eta^{\prime}$
and $\pi^{0}$ mesons rather than between the $\eta$ and $\eta^{\prime}$.

{\bf{Acknowledgements}}\\
We appreciate the work provided by the COSY operating team and thank them for the good cooperation and
for delivering the excellent proton beam. The research project was supported by the BMBF (06MS881I),
the Bilateral Cooperation between Germany and Poland represented by the Internationales B\"{u}ro DLR for 
the BMBF (PL-N-108-95), and by the FFE-grant (41266606 and 41266654)
from the Forschungszentrum J\"{u}lich. One of the authors (P.M.) acknowledges financial support from the 
Forschungszentrum J\"ulich and the Foundation for Polish Science.

\begin{table}
\caption[cross_etap]{
Total cross sections for the $pp \rightarrow pp\eta^{\prime}$ reaction
with respect to the excess energy in the center-of-mass system.
Only  statistical errors are quoted. In addition, there is an overall systematic uncertainty
of 15~$\%$ in the cross section and  0.44~MeV in energy. \\
}
\label{cross_etap}
\begin{tabular}[]{ccc}
Excess energy&Total cross section\\
\mbox{} [MeV] & \mbox{} [nb]\\
\hline
1.53~$\pm$~0.05 & 5.0~$\pm$~0.68  \\
2.11~$\pm$~0.20 & 6.9~$\pm$~1.4  \\
5.80~$\pm$~0.06 & 27.9~$\pm$~3.3  \\
7.57~$\pm$~0.07 & 43.5~$\pm$~4.3  \\
9.42~$\pm$~0.09 & 46.8~$\pm$~5.6  \\
10.98~$\pm$~0.12 & 67.4~$\pm$~8.2  \\
14.21~$\pm$~0.13 & 82.~$\pm$~13.  \\
23.64~$\pm$~0.20 & 140.~$\pm$~19.  \\
\end{tabular}
 \end{table}

\newlength{\ts}
\setlength{\ts}{\textwidth}
\addtolength{\ts}{-\columnsep}
\setlength{\ts}{0.5\ts}

\begin{figure}[t]
\epsfxsize=\ts
\vspace{2.5cm}
\centerline{\epsfig{figure=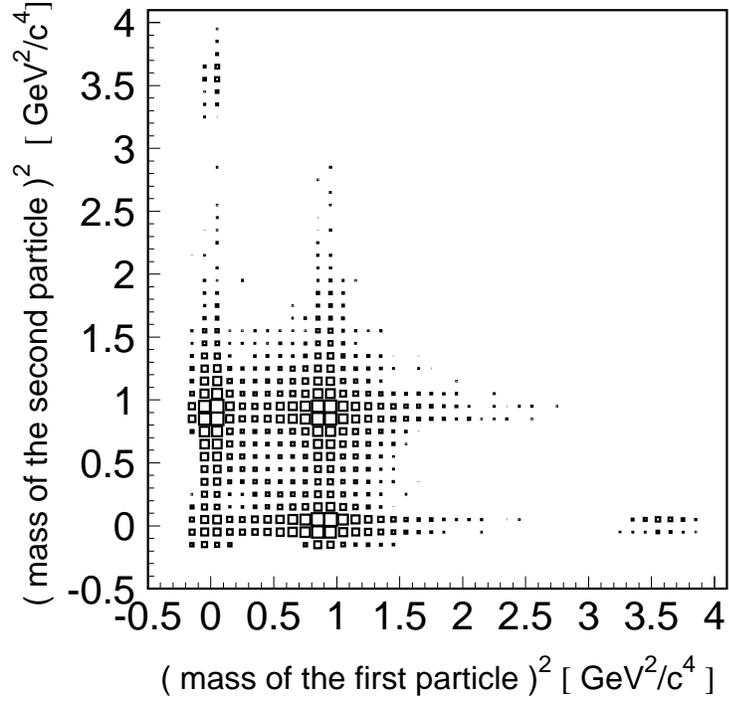,height=11.0cm,angle=0}}
\vspace{2.0cm}
\caption{
          Squared masses of two positively charged particles measured in coincidence.
          Pronounced peaks are to be recognized when two protons, proton and pion, two pions,
          or pion and deuteron were registered. Note that the number of events is
          shown in logarithmic scale.
        }
\label{invmass}
\end{figure}

\newpage

\newpage

\newpage

\begin{figure}[t]
\epsfxsize=\ts
\hspace{1.7cm}
\epsfig{figure=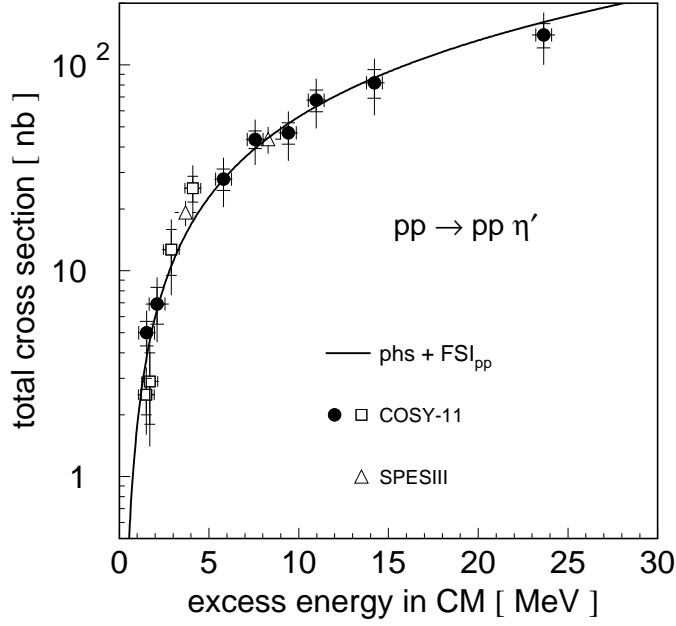,height=9.5cm,width=10.5cm,angle=0}
\vspace{2.0cm}
\caption{
         Total cross sections for the $pp\rightarrow pp\eta^{\prime}$ reaction as a function
         of the center-of-mass excess energy.
         Open squares and triangles are from references~\protect\cite{moskalprl}
         and~\protect\cite{hiboupl}, respectively.
         Filled circles  indicate the results of the 
         COSY~-~11 measurements reported in this letter. Corresponding numerical values are 
         given in table~\ref{cross_etap}. 
         Statistical and systematical errors are
         separated by dashes.
         The solid line shows the phase-space distribution with the inclusion of proton-proton
         strong and Coulomb interactions. 
        }
\label{etap-fsi}
\end{figure}

\end{document}